\documentclass[useAMS,usenatbib]{mn2e}
\pdfoutput=1
\usepackage{aas_macros}
\usepackage{epsfig}
\usepackage{amsmath}
\usepackage{amssymb}
\usepackage{leftidx}
\usepackage{natbib}
\usepackage{xcolor}
\usepackage{hyperref}
\hypersetup{colorlinks=true,citecolor=teal,linkcolor=violet}
\def\beq{\begin{equation}}
\def\eeq{\end{equation}}
\def\ba{\begin{eqnarray}}
\def\ea{\end{eqnarray}}
\def\bal{\begin{align}}
\def\eal{\end{align}}
\def\bxi{{\mbox{\boldmath $\xi$}}}
\def\bnab{{\mbox{\boldmath $\nabla$}}}

\begin{document}

\title[Giant Planet Tides] {Resonance locking as the source of rapid tidal migration in the Jupiter and Saturn moon systems}

\author[Fuller et al.]{
Jim Fuller$^{1,2}$\thanks{Email: jfuller@caltech.edu}, Jing Luan$^{3}$, and Eliot Quataert$^{3}$\\
\\$^1$TAPIR, Walter Burke Institute for Theoretical Physics, Mailcode 350-17, Caltech, Pasadena, CA 91125, USA
\\$^2$Kavli Institute for Theoretical Physics, Kohn Hall, University of California, Santa Barbara, CA 93106, USA
\\$^3$Astronomy Department and Theoretical Astrophysics Center, University of California at Berkeley, Berkeley, CA 94720-3411, USA}

\label{firstpage}
\maketitle

\begin{abstract}

The inner moons of Jupiter and Saturn migrate outwards due to tidal energy dissipation within the planets, the details of which remain poorly understood. We demonstrate that resonance locking between moons and internal oscillation modes of the planet can produce rapid tidal migration. Resonance locking arises due to the internal structural evolution of the planet and typically produces an outward migration rate comparable to the age of the solar system. Resonance locking predicts a similar migration timescale but a different effective tidal quality factor $Q$ governing the migration of each moon. It also predicts nearly constant migration timescales a function of semi-major axis, such that effective $Q$ values were larger in the past. Recent measurements of Jupiter and Saturn's moon systems find effective $Q$ values that are smaller than expected (and are different between moons), and which correspond to migration timescales of $\sim$10 Gyr. If confirmed, the measurements are broadly consistent with resonance locking as the dominant source of tidal dissipation in Jupiter and Saturn. Resonance locking also provides solutions to several problems posed by current measurements: it naturally explains the exceptionally small $Q$ governing Rhea's migration, it allows the large heating rate of Enceladus to be achieved in an equilibrium eccentricity configuration, and it resolves evolutionary problems arising from present-day migration/heating rates.

\end{abstract}


\section{Introduction}

Satellite systems around the outer planets exhibit rich dynamics that provide clues about the formation and evolution of our solar system. The orbits of the moons evolve due to tidal interactions with their host planets, and {\it every} moon system within our solar system shows strong evidence for significant tidal evolution within the lifetime of the solar system. Yet, in many cases, the origin of tidal energy dissipation necessary to produce the observed or inferred orbital migration remains poorly understood. 

The Jupiter and Saturn moon systems are particularly intriguing. In both systems, the planet spins faster than the moons orbit, such that tidal dissipation imparts angular momentum to the moons and they migrate outwards. The rich set of mean motion resonances (MMRs) between moons provide evidence for ongoing tidal migration because outward motion of inner moons produces convergent migration that allows moons to be captured into resonances (see discussion in \citealt{dermott:88,murray:99}). For instance, the 4:2:1 resonance of the orbits of Io:Europa:Ganymede indicates that Io has tidally migrated outwards by a significant fraction of its current semi-major axis, catching both Europa and Ganymede into MMRs during the process.

The tidal energy dissipation responsible for outward migration is often parameterized by a tidal quality factor $Q$ \citep{goldreich:66} that is difficult to calculate from first principles. Smaller values of $Q$ correspond to larger energy dissipation rates and shorter migration timescales. The actual value of $Q$ can be somewhat constrained by observed orbital architectures: very small values of $Q$ are implausible because they would imply the satellites formed inside their Roche radii, whereas very large values of $Q$ are unlikely because they would not allow for capture into MMRs within the lifetime of the solar system. In other words, the current orbital architectures mandate that outward migration timescales are within an order of magnitude of the age of the solar system.


In classical tidal theory, tidal energy dissipation occurs via the frictional damping of the equilibrium tide, which is defined as the tidal distortion that would be created by a stationary perturbing body. In this case, damping can arise from turbulent viscosity due to convective motions in a gaseous envelope \citep{goldreich:77} or by viscoelasticity in a solid core (\citealt{dermott:79,remus:12a,remus:12b,storch:14,remus:15}). However, tidal dissipation can also occur through the action of dynamical tides, which arise due to the excitation of waves and/or oscillation modes by the time-dependent gravitational force of the perturbing body. The dynamical tide may be composed of traveling gravity waves (e.g., \citealt{zahn:75,ioannou:93a,ioannou:93b}), tidally excited gravity modes (e.g., \citealt{lai:97,fuller:12b,burkart:12}, or tidally excited inertial modes/waves (e.g., \citealt{wu:05a,wu:05b,ogilvie:04,ogilvie:13,guenel:14,braviner:15,auclair:15,mathis:15}).

The notable feature of many previous works is that they struggle to produce tidal dissipation rates large enough to match those inferred for Jupiter and Saturn, given plausible internal structures of the planets. Models involving viscoelastic dissipation in the core or inertial waves in the envelope require the presence of a large core (solid in the former case, and either solid or more dense in the latter case) in order to produce significant energy dissipation. Although substantial cores likely do exist \citep{guillot:14}, water ice in giant planet cores is likely to be liquid \citep{wilson:12b}, and rocky materials (silicon and magnesium oxides) could be liquid or solid \citep{mazevet:15}. In both cases, the ice/rock is likely to be soluble in the surrounding hydrogen/helium \cite{wilson:12a}, and the cores may have substantially eroded (thereby erasing sharp density jumps) over the life of the solar system.  Moreover, viscoelastic models generally contain two free parameters (a core shear modulus and viscosity) which must be appropriately tuned in order to yield a tidal $Q$ compatible with constraints. 

Using remarkable astrometric observations spanning many decades, several recent works \citep{lainey:09,lainey:12,lainey:15} have provided the first (albeit uncertain) measurements of the outward migration rates of a few moons in the Jupiter and Saturn systems. These measurements indicate that current outward migration rates are much faster than predicted.  In fact, for a constant tidal $Q$, the current migration rates are incompatible with the contemporaneous formation of the moons and the planets, because the moons would have already migrated beyond their current positions over the lifetime of the solar system. The measurements also indicate that the effective tidal $Q$ is different for each moon, which cannot be explained by equilibrium tidal models that predict a nearly constant $Q$. 

In this work, we examine a mechanism known as resonance locking \citep{witte:99}, originally developed in stellar contexts, which produces accelerated orbital migration via dynamical tides. Previous planetary studies have neglected the fact that the internal structures of planets may evolve on timescales comparable to their age. Such structural evolution causes planetary oscillation mode frequencies to gradually change, allowing for resonance locking to occur. During a resonance lock, a planetary oscillation mode stays nearly resonant with the forcing produced by a moon, greatly enhancing tidal dissipation and naturally producing outward migration on a timescale comparable to the age of the solar system. In this paper, we examine the dynamics of resonance locking in giant planet moon systems, finding that resonance locking is likely to occur and can resolve many of the problems discussed above. Although there are substantial uncertainties in the planetary structure, evolution, and oscillation mode spectra of giant planets, resonance locking is insensitive to many of these details and yields robust predictions for outward moon migration on planetary evolution timescales.

The paper is organized as follows. In Section \ref{res}, we discuss tidal dissipation via resonances with oscillation modes and the process of resonance locking. Section \ref{orb} investigates orbital migration and MMRs resulting from resonance locking. We compare our theory with observations of the Jupiter and Saturn moon systems in Section \ref{jupsat}. Section \ref{discussion} provides discussion of our results, including implications for tidal heating and the orbital evolution of the moons. We summarize our findings in Section \ref{conclusions}. Details related to planetary oscillation modes, their ability to sustain resonance locks, and tidal heating can be found in the appendices.

\section{Tidal Migration via Resonances}
\label{res}

\subsection{Basic Idea}

Tidal dissipation and outward orbital migration of moons can be greatly enhanced by resonances between tidal forcing frequencies and discrete ``mode frequencies" associated with enhanced tidal dissipation. The mode frequencies may occur at gravity/inertial/Rossby mode frequencies, or they could correspond to frequencies at which inertial waves are focused onto attractors to create enhanced tidal dissipation \citep{rieutord:01,ogilvie:04,papaloizou:05,ogilvie:07,ivanov:07,goodman:09,rieutord:10,papaloizou:10,ivanov:10,ogilvie:12,auclair:15}. For the discussion that follows, the important characteristics of the mode frequencies are \\
1. The tidal energy dissipation rate at the mode frequencies is much larger than the surrounding ``continuum" dissipation rate.\\
2. The mode frequencies occupy narrow ranges in frequency space, i.e, adjacent resonances do not overlap. \\
3. The mode frequencies are determined by the internal structure of the tidally forced body and may evolve with time.\\

The outer layers of Jupiter and Saturn are composed of thick convective envelopes. These envelopes may allow for a dense spectrum of mode frequencies at which tidal dissipation is greatly enhanced by the action of inertial waves \citep{ogilvie:04}. For Saturn, recent observations of its rings \citep{hedman:13,hedman:14} have provided evidence that stable stratification exists deep within Saturn's interior \cite{fuller:14}. Stable stratification (or semi-convective layers) has also been advocated to exist based on Saturn's thermal evolution \citep{leconte:12,leconte:13}. Both stably stratified layers and semi-convective layers support gravity modes (g modes, see \citealt{belyaev:15}) that can enhance tidal dissipation.

To demonstrate the importance of resonances with mode frequencies, we calculate the frequencies and eigenfunctions of the g modes and fundamental modes (f modes) of the Saturn model presented in \cite{fuller:14}. We describe this process in more detail in Appendix \ref{modes}. We also calculate an approximate damping rate of each mode due to the turbulent viscosity acting within the convective envelope. Next, we calculate the energy dissipation rate due to the tidal excitation and turbulent damping of the modes. The efficiency of tidal dissipation can be expressed through the tidal migration timescale
\beq
\label{ttide}
t_{\rm tide} = - \frac{E_{\rm orb}}{\dot{E}_{\rm tide}} = \frac{a_{\rm m}}{\dot{a}_{\rm m,tide}} \, ,
\eeq
which describes the timescale on which a giant planet moon migrates outward due to tidal energy dissipation within the planet. Here, $\dot{E}_{\rm tide}$ is the orbital energy transferred to the moon by tides, and $E_{\rm orb} = -G M_{\rm p} M_{\rm m}/(2a_{\rm m})$ is the orbital energy, where $M_{\rm p}$ and $M_{\rm m}$ are the planetary and moon mass, respectively, and $a_{\rm m}$ is the semi-major axis of the moon. The efficiency of tidal dissipation can also be expressed in terms of the effective tidal quality factor $Q$, here defined as
\beq
\label{qdef}
Q \equiv 3 k_2 \frac{M_{\rm m}}{M_{\rm p}} \bigg(\frac{R_{\rm p}}{a_{\rm m}}\bigg)^5 \Omega_{\rm m} t_{\rm tide} \, .
\eeq
Here, $R_{\rm p}$ is radius of the planet, $k_2$ is its Love number, and $\Omega_{\rm m}$ is the moon's angular orbital frequency. 

\begin{figure}
\begin{center}
\includegraphics[scale=0.44]{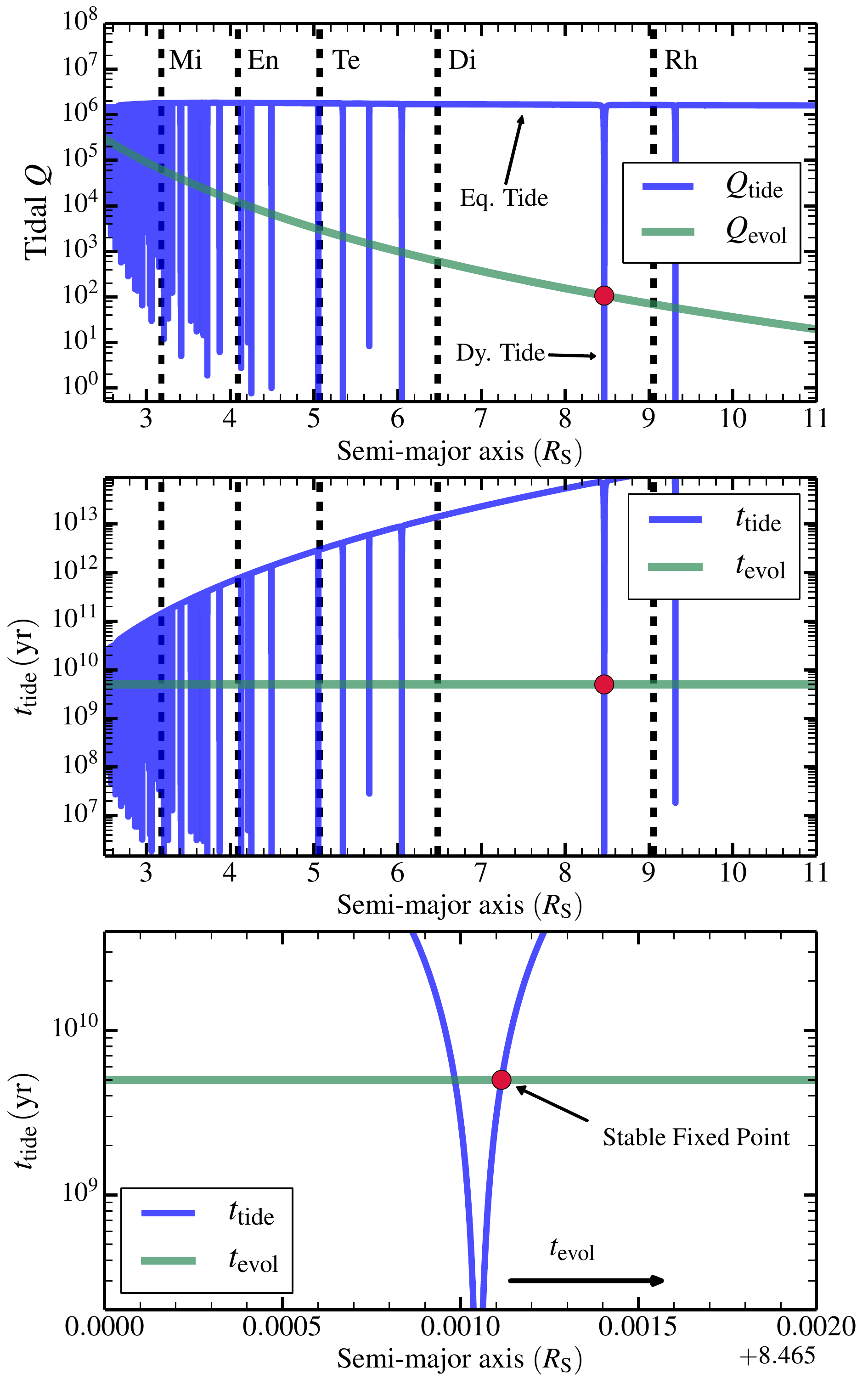}
\end{center} 
\caption{ \label{SatQmodes} {\bf Top:} Effective tidal quality factor $Q$ (equation \ref{qdef}) for Saturn tidally interacting with a moon with the mass of Tethys, as a function of semi-major axis for the Saturn model described in the text. Each sharp dip corresponds to a resonance with one of Saturn's g modes. The resonances are narrow and well-separated even at small semi-major axes, they merely appear to overlap because of the resolution of the plot. The smooth green curve $Q_{\rm evol}$ corresponds to the $Q$ required for a constant tidal migration time scale $t_{\rm evol} = T_\odot$. {\bf Middle:} Corresponding tidal migration time scale $t_{\rm tide}$ (equation \ref{ttide}), along with the constant time scale $t_{\rm evol}$. One location near a resonance where $t_{\rm tide}=t_{\rm evol}$ has been marked with a red circle. {\bf Bottom:} Zoom-in on a resonance. A moon in a resonance lock will remain at the stable fixed point, which moves outward on a timescale $\sim \! t_{\rm evol}$. }
\end{figure}

Figure \ref{SatQmodes} shows the values of $Q$ and $t_{\rm tide}$ as a function of semi-major axis due to tidal dissipation via oscillation modes of our Saturn model. The upper envelope of $t_{\rm tide}$ is set primarily by damping of the non-resonant tidally excited f modes of our model, and is essentially the effect of the equilibrium tide. The sharp dips in $t_{\rm tide}$ correspond to resonances with the g modes of our model.
Figure \ref{SatQmodes} also indicates the orbital distances of Saturn's five innermost major moons (Mimas, Enceladus, Tethys, Dione, and Rhea), which lie amongst resonances with Saturn's g modes.

At these resonances, the outward migration timescale may be reduced by several orders of magnitude. However, because the widths of the resonances are narrow, the average migration time scale is still quite long. A moon placed at a random semi-major axis would likely have a long tidal migration timescale, i.e., $t_{\rm tide} \gg T_\odot = 4.5 \, {\rm Gyr}$, with $T_\odot$ the age of the solar system. 

We emphasize that the existence of resonance-like features in Figure \ref{SatQmodes} is not highly dependent on the planetary model, mode damping rates, or the existence of g modes. Calculations that include the effects of inertial waves \citep{ogilvie:04} produce similar features in $t_{\rm tide}$: sharp dips at certain frequencies, with a longer frequency-averaged tidal migration time scale. The basic picture of enhanced tidal dissipation at resonances but weak dissipation away from resonances holds for many tidal models.

However, it is essential to realize that the mode frequencies are determined by the internal structure of the planet and therefore change on some time scale $t_{\rm evol}$. Jupiter and Saturn have dynamic interiors which are continuing to evolve due to processes such as cooling, helium sedimentation, and core dissolution which likely proceed on a time scale comparable to the age of the planets (see Section \ref{timescale} for more discussion on this issue). We thus expect $t_{\rm evol} \sim T_\odot$, to order of magnitude. The mode frequencies evolve on a similar time scale, and thus the locations of the resonant troughs in Figure \ref{SatQmodes} will sweep past the locations of the moons. This has two consequences. First, moons will pass through mode resonances even if their initial condition placed them far from resonance. Second, we show in the following section that moons may lock into resonance with a mode as it sweeps past, allowing it to ``surf" the resonance and migrate outward at a greatly reduced timescale of order $\sim \! T_\odot$.

\subsection{Resonance Locking}
\label{reslocking}

The structures of planets evolve with time, as do the frequencies of their oscillation modes. We define the time scale on which the angular frequency $\omega_\alpha$ of an oscillation mode changes as
\beq
\label{omalphadot}
\dot{\omega}_\alpha = \frac{\omega_\alpha}{t_\alpha} \, .
\eeq
In general, the time scale $t_\alpha$ is comparable to the stellar/planetary evolution time scale. For Jupiter and Saturn, we leave $t_\alpha$ as a free parameter, and make informed estimates of it in Section \ref{timescale}. The effect of changing mode frequencies can drastically alter tidal evolution timescales by allowing for a process called resonance locking, originally examined by \citealt{witte:99,witte:01} (see also \citealt{fuller:12b,burkart:12,burkart:14}). During resonance locking, the coupled evolution of the mode frequency and the orbital frequency of the perturbing body proceeds such that the perturber remains near resonance with the oscillation mode. 

The dynamics of resonance locking can be qualitatively understood from the bottom panel of Figure \ref{SatQmodes}. Consider a moon located at the stable fixed point, where its outward migration timescale is equal to that at which the resonant location moves outward. If its orbit is perturbed inward (toward resonance) tidal dissipation will increase, and the moon will be pushed back outward toward the fixed point. If the moon's orbit is perturbed outward (away from resonance) the moon's outward migration rate will decrease,
and the resonant location will move outward and catch up with the moon. Thus, the moon can ``ride the tide" and stably migrate outward with the location of a resonance. While locked in resonance, the tidal migration timescale is drastically reduced compared to its value away from resonances, and the moon's orbit evolves on a time scale comparable to $t_\alpha$. 

In what follows, we make some simplifying assumptions. Because the moon masses are very small ($M_{\rm m} \ll M_{\rm p}$), we may safely neglect the backreaction on (i.e., the spin-down of) the planet. Moreover, for our purposes we can approximate the moons' orbits as circular ($e = 0$) and aligned with the spin axis of the planet ($i = 0$). We shall account for MMRs between moons in Section \ref{moonres}. In this paper, we use the convention that the mode displacement in the rotating frame of the planet is proportional to $e^{i(\omega_{\alpha} t + m \phi)}$, where $m$ is the azimuthal number of the mode, such that modes with positive frequency and positive $m$ are retrograde modes in the planet's frame.

In the outer planet moon systems, an oscillation mode near resonance with a moon has an angular frequency (measured in the planet's rotating frame)
\beq
\label{res1}
\omega_\alpha \simeq  \omega_{\rm f} = m ( \Omega_{\rm p} - \Omega_{\rm m} )  \,
\eeq 
where $\Omega_{\rm p}$ is the angular spin frequency of the planet, and $\omega_{\rm f}$ is the forcing frequency of the moon measured in the planet's rotating frame.  Because the planet rotates faster than the moons orbit, the resonant modes are retrograde in the rotating frame of the planet, but prograde in the inertial frame. Resonant modes also have $\omega_\alpha \! < \! 2 \Omega_{\rm p}$ for $m \!= \! 2$, and may therefore lie in the sub-inertial range where mode properties become more complex, which we discuss more in Appendix \ref{modes}. For the moons considered in this work, only Mimas has $\omega_{\rm f} \! < \! 2 \Omega_{\rm p}$ for $m \geq 3$, and so most resonant $m \geq 3$ modes do not lie in the sub-inertial regime. Typical mode periods are $P_\alpha = 2 \pi/\omega_\alpha \sim 3 \times 10^4 \, {\rm s}$.


\subsection{Resonance Locking with Inertial Waves}
\label{inertial}

Resonance locking with inertial waves may be difficult to achieve. Consider a ``mode" frequency at which there is enhanced tidal dissipation due to inertial waves. We assume this mode frequency (measured in Saturn's rotating frame) scales with Saturn's rotation rate such that
\beq
\label{ominertial}
\omega_\alpha = c \Omega_p \,
\eeq
where $c$ is a constant and $|c| \! < \! 2$ (see discussion in \citealt{ogilvie:04}). For a retrograde mode that can resonate with the moons, the frequency in the inertial frame is $\sigma_\alpha = (c-m) \Omega_{\rm p}$, where $m \! > \! 0$ is the azimuthal number of the mode. Resonance occurs when
\beq
\label{lock}
-m \Omega_{\rm m} = (c-m) \Omega_{\rm p} \, ,
\eeq
and a resonance lock requires
\beq
\label{lock1}
m \dot{\Omega}_{\rm m} = (m-c) \dot{\Omega}_{\rm p} \, 
\eeq
or equivalently
\beq
\label{lock2}
\frac{\dot{\Omega}_{\rm m}}{\Omega_{\rm m}} = \frac{\dot{\Omega}_{\rm p}}{\Omega_{\rm p}} \, .
\eeq
Since $\dot{\Omega}_{\rm m}<0$ as a moon migrates outward, resonance locking requires $\dot{\Omega}_{\rm p}<0$, i.e., it requires the planet to be spinning down. Therefore, unless the value of $c$ changes due to internal structural evolution, resonance locking with inertial waves cannot occur due to planetary contraction and spin-up. However, we note that $c$ is a function of the rotation frequency (see, e.g., Figure 19 of \citealt{papaloizou:10}) and is not expected to be exactly constant. Moreover, in the realistic case of an evolving density profile, the value of $c$ will change because the frequencies of inertial modes depend on the density profile (see, e.g., \citealt{ivanov:07,ivanov:10}).

No studies of inertial waves in evolving planets have been performed. We encourage such studies to determine how the value of $c$ will change in an evolving planet. If $\dot{c} > 0$, resonance locking with inertial waves may be possible and would provide an avenue for enhanced tidal dissipation in generic models of giant planets.

\section{Tidal Migration of Resonantly Locked Moons}
\label{orb}

While a moon is caught in a resonance lock, the resulting tidal dynamics and outward migration rate are simple to calculate. Differentiating the resonance criterion of equation \ref{res1} with respect to time leads to the locking criterion
\beq
\label{reslock}
 m ( {\dot \Omega}_{\rm p} - {\dot \Omega}_{\rm m} ) \simeq {\dot \omega}_\alpha \, .
\eeq
We recall the definition of the mode evolution timescale $t_{\alpha} = \omega_\alpha/\dot{\omega}_\alpha$, and similarly define the planetary spin evolution timescale $t_{\rm p} = \Omega_{\rm p}/\dot{\Omega}_{\rm p}$. Equation \ref{reslock} becomes
\beq
\label{omdot}
\dot{\Omega}_{\rm m} = \frac{\Omega_{\rm p}}{t_{\rm p}} - \frac{\omega_\alpha}{m t_\alpha} \, ,
\eeq
and hence the moon's semi-major axis evolves as 
\beq
\label{adot}
\frac{\dot{a}_{\rm m}}{a_{\rm m}} = \frac{2}{3} \bigg[ \frac{\omega_\alpha}{m \Omega_{\rm m} t_\alpha} - \frac{\Omega_{\rm p}}{\Omega_{\rm m} t_{\rm p}}  \bigg] \, .
\eeq
This can be rewritten
\beq
\label{ttidein}
\frac{1}{t_{\rm tide}} = \frac{2}{3} \bigg[ \frac{\Omega_{\rm p}}{\Omega_{\rm m}} \bigg(\frac{1}{t_\alpha} - \frac{1}{t_{\rm p}}\bigg) - \frac{1}{t_\alpha}  \bigg] \, .
\eeq
Therefore, during resonance locking, the orbital migration rate is determined by the evolutionary timescale of the {\it planet}. Note that outward migration requires $0 \! < \! t_{\alpha} \! < \! t_{\rm p}$, which is less restrictive than the condition for inertial waves discussed in Section \ref{inertial}.

Using equations \ref{ttide} and \ref{qdef}, the effective value of $Q$ during a resonance lock is
\beq
\label{q}
Q_{\rm ResLock} = \frac{9 k_2}{2} \frac{M_{\rm m}}{M_{\rm p}} \bigg(\frac{R}{a}\bigg)^{\!5} \bigg[ \frac{\omega_\alpha}{m \Omega_{\rm m}^2 t_\alpha} - \frac{\Omega_{\rm p}}{\Omega_{\rm m}^2 t_{\rm p}}  \bigg]^{-1} \, .
\eeq
In addition to the physical parameters of the system (mass, radius, etc.), the primary factor controlling $Q_{\rm ResLock}$ of a moon are the evolution time scales $t_\alpha$ and $t_{\rm p}$.  
According to the resonance locking hypothesis, $Q$ is not a fundamental property of the planet. Instead, a resonantly locked moon migrates outward on a timescale comparable to $t_\alpha$ and $t_{\rm p}$, which {\it are} fundamental properties of the planet in the sense that they are determined by the planetary evolution.

\subsection{Accounting for Mean-Motion Resonances}
\label{moonres}

As moons migrate outward, they may become caught in MMRs with outer moons. If trapped into MMR with an outer moon, an inner moon may still move outward via resonance locking on a time scale $t_\alpha$, such that both moons migrate outward on this time scale. The resonance with the outer moon effectively increases the inertia of the inner moon, such that a smaller $Q$ is required to push it outward on the timescale $t_\alpha$. To accommodate the added inertia, the inner moon must move deeper into resonance with the planetary oscillation mode to remain resonantly locked.

Consider two moons migrating in MMR, with the inner moon denoted by subscript 1 and the outer moon denoted by subscript 2. In a first order MMR (ignoring for simplicity the splitting of resonances due to precession/regression), the moons' orbital frequencies maintain the relation
\beq
\label{j}
j \, \Omega_2 = (j-1) \Omega_1 \, .
\eeq
It follows that in an MMR,
\beq
\label{a12}
\frac{\dot{a}_1}{a_1} = \frac{\dot{a}_2}{a_2} \, .
\eeq
Now, the total rate of change of angular momentum of the orbits of both moons is
\begin{align}
\label{jdot}
\dot{J} &= \dot{J}_1 + \dot{J}_2 \nonumber \\
&= \frac{1}{2} M_1 \sqrt{GM_{\rm p}a_1} \frac{\dot{a}_1}{a_1} + \frac{1}{2} M_2 \sqrt{GM_{\rm p} a_2} \frac{\dot{a}_2}{a_2} \nonumber \\
&= \frac{1}{2} \frac{\dot{a}_1}{a_1} \big( J_1 + J_2 \big) \, .
\end{align}
The last equality results from the condition \ref{a12}.

The rate at which the orbital energy is increasing due to tides raised in the planet is 
\begin{align}
\label{edottide}
\dot{E}_{\rm tide} &= \dot{E}_{1,{\rm tide}} + \dot{E}_{2,{\rm tide}} \nonumber \\
&= \Omega_1 \dot{J}_{1,{\rm tide}} + \Omega_2 \dot{J}_{2,{\rm tide}} \, .
\end{align}
This is not equal to the rate at which orbital energy changes because orbital energy may be tidally dissipated as heat within the moons if their orbits become eccentric. Additionally, the relation $\dot{E}_{1,{\rm tide}}/E_1 = \dot{a}_1/a_1$ no longer holds.

Let us consider the limiting case in which all the tidal dissipation within the planet is caused by the inner moon. Then we have $\dot{E}_{\rm tide} = \dot{E}_{1,{\rm tide}} = \Omega_1 \dot{J}_{1,{\rm tide}} = \Omega_1 \dot{J}$. Using the relation above, the definition of $Q$ from equation \ref{qdef}, and the last line of equation \ref{jdot}, we have
\beq
\label{eq}
\frac{\dot{E}_{1,{\rm tide}}}{E_1} = - \frac{9}{2} \frac{k_2}{Q_1} \frac{M_1}{M_{\rm p}} \bigg(\frac{R_{\rm p}}{a_1}\bigg)^{\! 5} \Omega_1 \, = \frac{1}{2} \frac{\dot{a}_1}{a_1} \frac{ \Omega_1 \big( J_1 + J_2 \big) }{E_1} \, .
\eeq
Since the moons' orbits are nearly circular, then $E_1 \simeq - \Omega_1 J_1/2$. Moreover, if the inner moon remains caught in the resonance lock, equation \ref{adot} still holds. Substituting these above, we find 
\beq
\label{q2}
Q_{1,{\rm min}} = \frac{9 k_2}{2} \frac{M_1}{M_{\rm p}} \bigg(\frac{R_{\rm p}}{a_1}\bigg)^{\!5} \bigg[ \frac{\omega_\alpha}{m \Omega_1^2 t_\alpha} - \frac{\Omega_{\rm p}}{\Omega_1^2 t_{\rm p}} \bigg]^{-1} \bigg[1 + \frac{J_2}{J_1} \bigg]^{-1} \, .
\eeq
This corresponds to a minimum value of $Q_1$ because it assumes no tides from the outer moon. Equation \ref{q}, in turn, is a maximum tidal $Q$ for the inner moon caught in MMR. The two are related by a factor $(1 + J_2/J_1)$ which accounts for the extra dissipation needed to drive the exterior moon outward.

Additionally, there is a minimum possible $Q_2$ for the outer moon to remain in MMR, because it will escape from resonance if it migrates outward faster than the inner moon. This minimum $Q_2$ is found by setting $\dot{a}_1/a_1 = \dot{a}_2/a_2$, and letting each moon migrate outward via its own tides raised in the planet. In this case, we have 
\beq
\label{q12}
Q_{2,{\rm min}} = Q_1 \frac{M_2}{M_1} \bigg( \frac{a_1}{a_2} \bigg)^{\! 5} \frac{j-1}{j} \, ,
\eeq
with $Q_1$ evaluated from equation \ref{q}. The MMR requires $a_1 = \big[(j-1)/j\big]^{2/3} a_2$. Then
\beq
\label{q122}
Q_{2,{\rm min}} = Q_1 \frac{M_2}{M_1} \bigg(\frac{j-1}{j}\bigg)^{\! 13/3} \, .
\eeq
Since the outer moon may be pushed out solely by the resonance with the inner moon, there is no maximum possible value of $Q_2$ to remain in MMR.

\section{Comparison with Jupiter and Saturn Systems}
\label{jupsat}

\begin{figure}
\begin{center}
\includegraphics[scale=0.44]{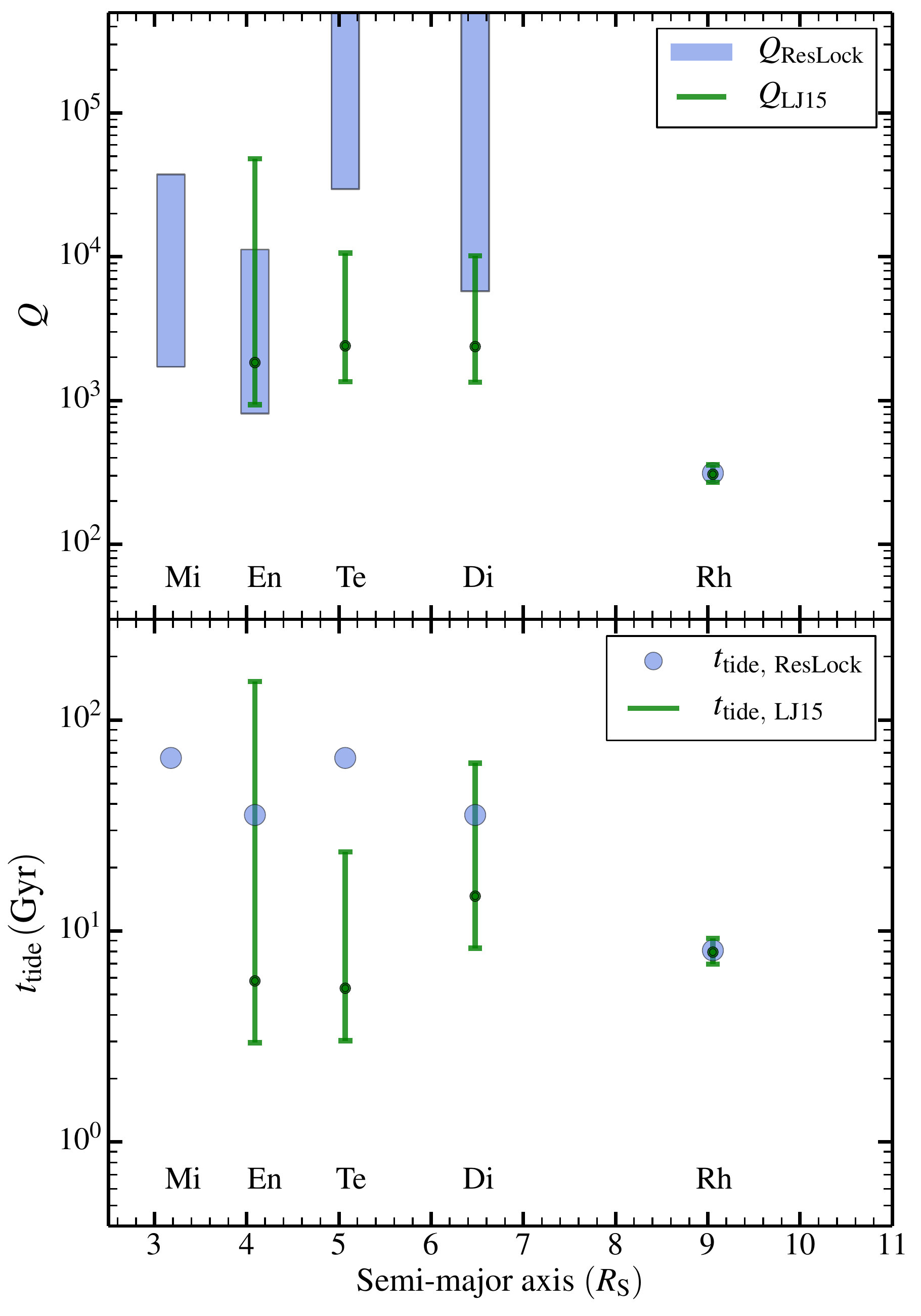}
\end{center} 
\caption{ \label{SatQ} 
{\bf Top:} Effective tidal quality factors $Q$ for Saturn interacting with its inner moons. The green points are the measurements of \citealt{lainey:15}. The blue boxes are the predicted values of $Q$ for $t_\alpha = 50 \, {\rm Gyr}$ and $t_{\rm p} = \infty$ in the resonance locking theory. For Mimas, the lower bound on the predicted $Q_{\rm ResLock}$ is that required to maintain the resonance lock and push out Tethys, assuming all the tidal dissipation within Saturn is produced by a resonance lock with Mimas (equation \ref{q2}). The upper bound on $Q_{\rm ResLock}$ for Mimas is that required to maintain the resonance lock, but with Tethys migrating outward due to its own tides (equation \ref{q}). For Tethys, the lower bound in $Q_{\rm ResLock}$ is the minimum value of $Q$ such that it remains in mean motion resonance with Mimas (equation \ref{q122}). There is no upper bound to $Q$ for Tethys because it may be pushed outwards solely by Mimas. The Enceladus-Dione system behaves similarly. Rhea is not in a mean-motion resonance, allowing for a precise prediction. {\bf Bottom:} Corresponding outward migration time scale $t_{\rm tide}$. The observed values of $t_{\rm tide}$ are calculated from equation \ref{qdef} using the measured values of $Q$ and do not take mean-motion resonances into account (see text). The data is consistent with $t_{\rm tide} \approx 10 \, {\rm Gyr}$ for all moons.}
\end{figure}

We now apply our theories to the Jovian and Saturnian moon systems, and compare with the recent measurements of $Q$ from \cite{lainey:09,lainey:12,lainey:15}.

\subsection{Saturn}


Figure \ref{SatQ} shows the effective values of $Q$ measured by \cite{lainey:15} for the tidal interaction between Saturn and its inner moons. Although the measured value of $Q$ is similar for Saturn's interaction with Enceladus, Tethys, and Dione, it is roughly one order of magnitude smaller for Rhea. The much smaller $Q$ for Rhea's migration cannot be explained by any model of equilibrium tidal energy dissipation, and can only be accounted for by models including dynamical tides. Interestingly, the bottom panel of Figure \ref{SatQ} shows that the corresponding outward migration timescale is similar for each moon, with $t_{\rm tide} \sim 10 \, {\rm Gyr}$.

We have computed the measured values of $t_{\rm tide}$ from the measured values of $Q$ using equation \ref{qdef}. This equation does not take MMRs into account, which can change the actual migration timescale. For Enceladus, the measured value of $t_{\rm tide}$ should be regarded as a lower limit, since its migration may be slowed by outer moons. For Tethys and Dione, the value of $t_{\rm tide}$ is an upper limit, since they may be pushed outward by inner moons. Unfortunately, all of these timescales are dependent on one another. They also depend on inward migration rates due to eccentricity damping, which in turn depend on the values of $k_2$ and $Q$ for tidal effects within each moon, which are not constrained by \cite{lainey:15}. 

With these complications in mind, we also plot in Figure \ref{SatQ} the values $Q$ and $t_{\rm tide}$ expected for the resonance locking scenario. We have set $t_{\rm \alpha} = 50 \, {\rm Gyr}$ in equation \ref{q} such that the value of $Q$ from resonance locking roughly matches the observed value for Rhea, which offers the best chance for comparison due to its lack of MMRs. For simplicity, we adopt the limit of no planetary spin-up, $t_{\rm p}\rightarrow\infty$. The remarkable result of this exercise is that resonance locking naturally produces a very low effective $Q$ for Rhea even for a realistic (and perhaps somewhat slow, $t_{\rm \alpha} \sim 10 \, T_\odot$) mode evolution timescale. 

The predicted values of $Q$ and $t_{\rm tide}$ from resonance locking for the inner moons (Enceladus, Tethys, and Dione) are similar to but slightly larger than the measurements of \cite{lainey:15}. The predicted values for Tethys are most discrepant, and are incompatible with the value $t_\alpha = 50 \, {\rm Gyr}$ that fits for Rhea. This may indicate that the inner moons migrate outward due to viscoelastic dissipation in the core as advocated in \cite{lainey:15}. However, the fact that $t_{\rm tide} \! \sim \! 10 \, {\rm Gyr}$ for each of these moons indicates that dynamical tides and/or resonance locking could still be occurring. Since $t_\alpha$ is not expected to be the same for each oscillation mode, it is possible that the effective $t_\alpha$ driving resonance locking of the innermost moons is smaller by a factor of a few compared to the best fit value for Rhea.

We caution that the MMRs between these moons may complicate both the measurements and their interpretation. For instance, the eccentricity of Enceladus is excited by the MMR with Dione, allowing inward migration due to tidal dissipation within Enceladus. The current outward migration of both Enceladus and Dione may not represent an equilibrium or time-averaged migration rate (see e.g., \citealt{meyer:07}), depending on the dynamics of tidal dissipation within Enceladus. Moreover, the measured values of $Q$ in \cite{lainey:15} for Enceladus, Tethys, and Dione were all dependent on the dissipation within Enceladus, and so we believe these measurements should be interpreted with caution. Similarly, the migration of Tethys is affected by its MMR with Mimas, for which the outward migration was not constrained by \cite{lainey:15}, and which may influence the measured value of $Q$ for the other moons.

\subsection{Jupiter}

\begin{figure}
\begin{center}
\includegraphics[scale=0.44]{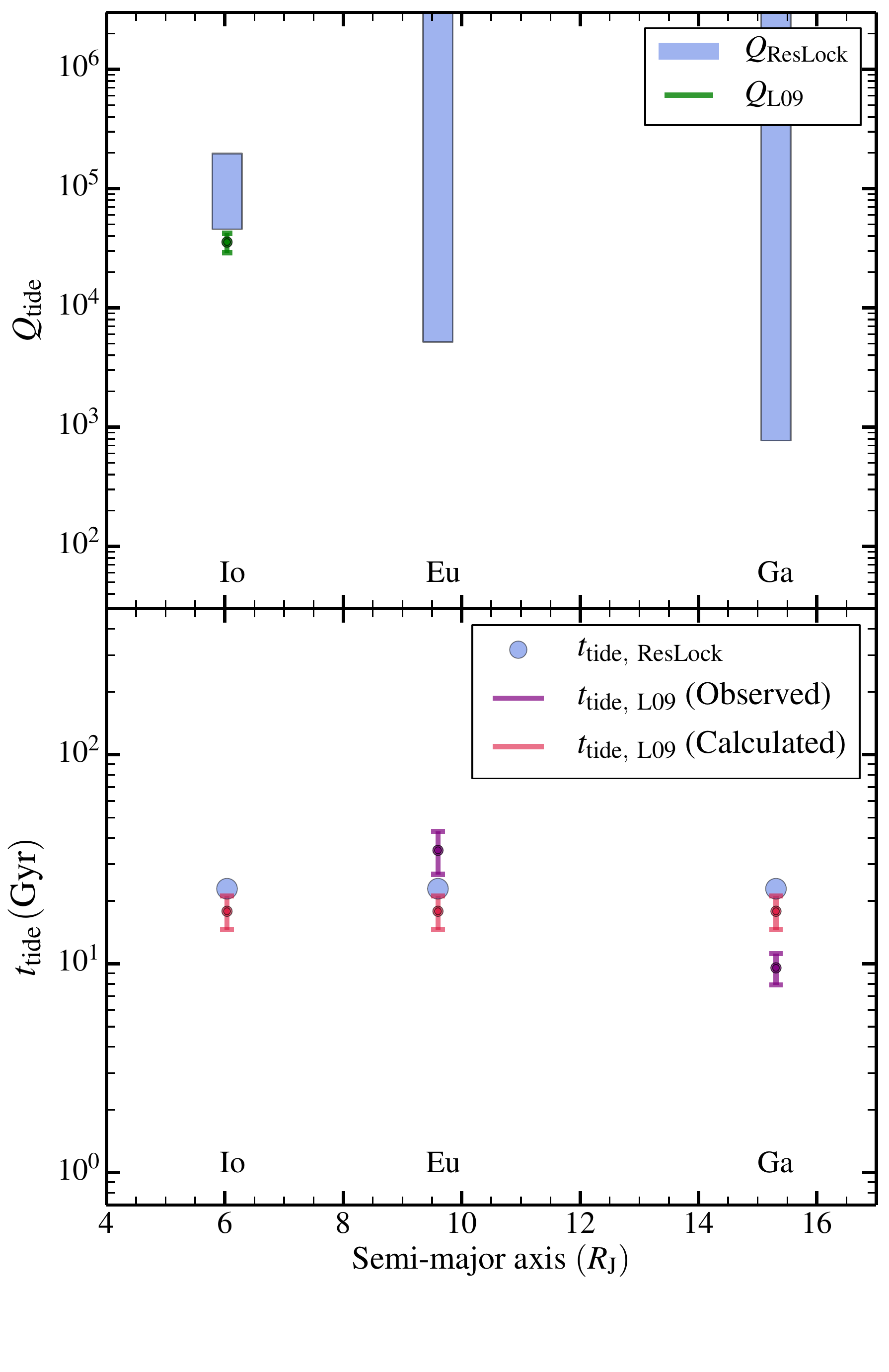}
\end{center} 
\caption{ \label{JupQ} 
{\bf Top:} Effective tidal quality factors $Q$ for Jupiter interacting with its inner moons. The green point is the measurement of \citealt{lainey:09}. The blue boxes are the predicted values of $Q$ using the same $t_\alpha = 50 \, {\rm Gyr}$ as in Saturn (see Figure \ref{SatQ}). {\bf Bottom:} Corresponding outward migration time scale $t_{\rm tide}$. Observed points are taken from \citealt{lainey:09}, where the measured inward migration of Io has $t_{\rm tide} \! < \! 0$ and is not shown. We also plot the calculated outward migration timescale of the Io:Europa:Ganymede chain from the measured value of $Q$ for the Io-Jupiter interaction, assuming zero tidal dissipation produced by Europa and Ganymede.}
\end{figure}

Figure \ref{JupQ} shows the predicted and measured \citep{lainey:09} values of $Q$ and $t_{\rm tide}$ for the moons of Jupiter. We have again used $t_\alpha = 50 \, {\rm Gyr}$, although the appropriate value of $t_\alpha$ could be different for Jupiter. The measured $t_{\rm tide}$ for Io was negative, with the interpretation that the instantaneous migration of Io is inward due to eccentricity damping. This point does not appear on the plot, and likely does not represent the long-term migration rate of Io. We have also plotted a value of $t_{\rm tide}$ calculated from the measured $Q$ for Io, by assuming it drives the migration of Europa and Ganymede in the current MMR. We caution against making a very thorough comparison with the data, as it is unclear whether the measured value of $Q$ for Io will be modified with updated measurements, as was the case in the Saturnian system.

Nonetheless, we note that our predicted $Q$ for Io falls very close to the measured value, especially if we take the lower bound on the predicted $Q$ corresponding to Io driving the outward migration of Europa and Ganymede. This correspondence can also be seen by the proximity of the predicted value of $t_{\rm tide}$ with the calculated $t_{\rm tide}$ which may more accurately characterize the long-term migration than the instantaneous observed $t_{\rm tide}$. This entails an outward migration timescale of $t_{\rm tide} \sim 20 \, {\rm Gyr}$ for Io, Europa, and Ganymede. In this case, we expect the effective tidal $Q$ for tidal dissipation in Jupiter caused by Europa to be quite large, $Q \! \gtrsim \! 10^4$, although we expect Europa to be migrating outward at $t_{\rm tide} \sim 20 \, {\rm Gyr}$ due to its MMR with Io.

\section{Discussion}
\label{discussion}

\subsection{Relation with Previous Tidal Theories}

The measurements of \cite{lainey:15} clearly rule out the common assumption of a constant value of $Q$ governing the outward tidal migration of Saturn's moons. The majority of previous literature investigating the subject assumed a constant $Q$, and must now be interpreted with caution.

Several works (e.g., \citealt{remus:12a,guenel:14,lainey:15}) have sought to explain the tidal $Q\sim 2000$ measured for Saturn due to forcing by Enceladus, Tethys and Rhea via viscoelastic dissipation within a solid core. However, this conclusion generates problems for the orbital evolution of the moons (see Section \ref{orbits}), forcing one to accept that the moons formed billions of years after the rest of the solar system, or that $Q$ was much larger (for no obvious reason) in the past. Additionally, these theories cannot explain the small $Q$ of Saturn due to forcing by Rhea.

Resonance locking can resolve many of these puzzles. It accounts for varying values of $Q$ by positing a nearly constant value of $t_\alpha$ which governs the value of $Q$ through equation \ref{q}. Unlike the value of $Q$ in equilibrium tidal theory that can be very difficult to compute from first principles (and frequently yields predictions orders of magnitude too large), the value of $t_\alpha$ can be calculated based on a thermal evolution model of a planet. The natural expectation is that $t_\alpha$ is comparable to the age of the solar system, yielding outward migration timescales of similar magnitude, as observed. 

Although other dynamical tidal theories (e.g., \citealt{ogilvie:04,ogilvie:12,auclair:15}) can produce low and varying values of $Q$, they suffer from two problems. First, the widths of the resonances at which strong tidal dissipation occurs are somewhat narrow (especially for small core radii), and it is not clear whether we expect to find any moons within these resonances. Second, at an arbitrary orbital frequency there is no reason to expect $t_{\rm tide}$ to be comparable to the age of the solar system. Resonance locking solves both these problems because moons can get caught in resonance locks that could last for billions of years, making it likely to observe a moon in a state of rapid outward migration. Moreover, the value of $t_{\rm tide}$ is naturally expected to be comparable to the age of the solar system.

Our purpose here is not at all to dismiss tidal theories based on dissipation of inertial waves. Instead, we advocate that these theories naturally reproduce the observations only if planetary evolution is included in the long-term behavior of the system. The evolution of the planet causes the locations of tidal resonances to migrate, allowing for resonance locks with the moons such that they migrate at a similar rate. In this picture, the precise frequencies and strengths of tidal resonances with inertial waves or g modes is not important, nor is their frequency-averaged dissipation rate. All that matters is the mere existence of such resonances, and the rate at which these resonant frequencies evolve (see equation \ref{ttidein}).

\subsection{Evolutionary Timescales}
\label{timescale}

In the resonance locking scenario, the outward migration timescale of moons is set by the planetary evolution timescale $t_{\alpha}$ in equation \ref{omalphadot}. Predicting this timescale is not simple, as it depends on internal structural evolution timescales, which are poorly constrained. However, we can place some rough constraints. First, we expect $t_\alpha$ to be comparable to or longer than the age of the solar system $T_{\odot}$. Any process that occurs on a shorter timescale has already occurred, or has slowed down to timescales of $\sim \! T_{\odot}$. 

Second, we can estimate an upper limit from the planet's thermal emission, which is generated through the release of gravitational energy. The intrinsic power radiated by Saturn is $L_{\rm Sa} \simeq \ 8.6 \times 10^{23} \, {\rm erg} \, {\rm s}^{-1}$ \cite{guillot:14}, likely generated via gravitational energy released through helium rain out. The corresponding Kelvin-Helmholtz time of Saturn is 
\beq
\label{Tsa}
T_{\rm Sa} = \frac{G M_{\rm Sa}^2}{R_{\rm Sa} L_{\rm Sa}} \approx 100 \, {\rm Gyr} \, .
\eeq
Therefore, we expect the frequencies of oscillation modes in Saturn to be changing on timescales $4.5 \, {\rm Gyr} \lesssim t_{\alpha} \lesssim 100 \, {\rm Gyr}$. Cooling timescales $t_{\rm cool} = T_{\rm ef}/(d T_{\rm ef}/dt) \sim 25 \, {\rm Gyr}$ found by \cite{fortney:11} and \cite{leconte:13}are within this range, and are comparable with the best fit timescale $t_\alpha = 50 \, {\rm Gyr}$ for Rhea in Figure \ref{SatQ}.

Importantly, we also expect that the mode frequencies increase with time (as measured in the rotating frame), as required for a stable resonance lock with Saturn's moons. Mode frequencies determined by internal structure typically scale with the planet's dynamical frequency, which increases due to gravitational contraction. Moreover, ongoing helium sedimentation or core erosion that builds a stably stratified layer (found to be present via Saturn ring seismology, see \citealt{fuller:14}) will cause g mode frequencies to increase (see Appendix \ref{resonancelocks}).

When a resonance lock is active, the outward migration timescale is 
\beq
\label{ttiderl}
t_{\rm tide} \approx \frac{3}{2} \frac{ \Omega_{\rm m}}{ \Omega_{\rm p} - \Omega_{\rm m}} t_{\alpha} \, ,
\eeq
in the limit $t_{\rm p} \rightarrow \infty$. For Mimas, $t_{\rm tide} \sim 1.5 \, t_\alpha$, but for Rhea, $t_{\rm tide} \sim 0.16 \, t_{\alpha}$. A resonance lock cannot persist indefinitely, as it requires $t_{\rm tide} \rightarrow 0$ as $\Omega_{\rm m} \rightarrow 0$. As a moon migrates outward, the resonance lock will eventually break when the required mode amplitude becomes too large. This can occur due to non-linear effects, or because the stable fixed point disappears when it reaches the center of the resonant trough in Figure \ref{SatQmodes} (i.e., the resonance saturates). For the g mode resonances shown in Figure \ref{SatQmodes}, the modes are linear in the sense that fluid displacements are orders of magnitude smaller than their wavelengths. Resonant saturation could occur if the mode damping rates are a few orders of magnitude larger than those calculated in Appendix \ref{damp}. We find this unlikely, given the long mode lifetimes of the f modes resonating with Saturn's rings \citep{hedman:13}.

\begin{figure}
\begin{center}
\includegraphics[scale=0.44]{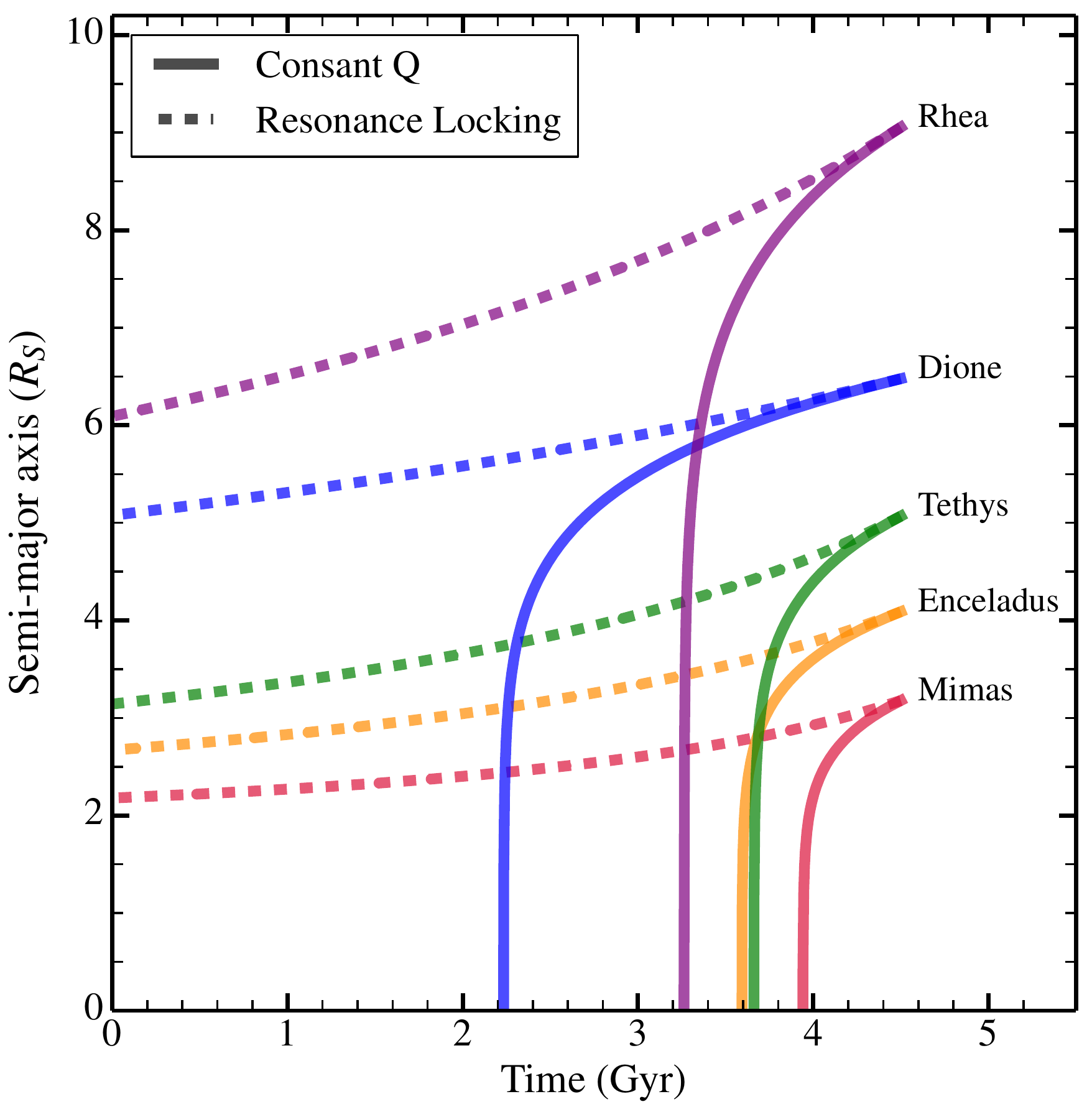}
\end{center} 
\caption{ \label{SatQevol} 
Evolution of moon semi-major axes in the Saturn system, calculated by integrating orbital evolution equations backward from the present era at $t \simeq 4.5 \, {\rm Gyr}$. Solid lines are calculated with a constant tidal $Q$ theory, using measured values of $Q$ \citep{lainey:15}. We have used $Q=2000$ for Mimas. Dashed lines correspond to the resonance locking theory, with $t_\alpha = 50 \, {\rm Gyr}$ (approximately consistent with observations, see text). These orbital evolutions do not take mean-motion resonances into account and are only meant to illustrate the qualitative behavior of the different tidal theories. The constant tidal $Q$ theory is incompatible with coeval formation of Saturn and its moons, whereas resonance locking allows for coeval formation.
}
\end{figure}

\subsection{Orbital Evolution of the Moons}
\label{orbits}

Resonance locking yields qualitatively different orbital evolution compared to any constant $Q$ theory. For a constant $Q$, the tidal migration timescale for a moon of a given mass is a very strong function of semi-major axis (see Figure \ref{SatQmodes}), such that $t_{\rm tide} \propto a^{13/2}$ (see equation \ref{qdef}). This has created problems for conventional tidal theories, because it implies that outward migration rates were much faster in the past, thereby requiring large values of $Q$ (and correspondingly slow present-day outward migration) in order for the moons to have migrated to their current positions if they formed coevally with Saturn. However, the measured values of $Q$ are much smaller than the lower limits described by \cite{peale:80}, implying such rapid past migration (in a constant $Q$ scenario) that the moons could not have formed at the same time as Saturn. Studies which have assumed constant values of $Q$ for Saturn in order to constrain the orbital evolution history (e.g., \citealt{peale:80,meyer:07,meyer:08,zhang:09}) should be interpreted with caution.

Figure \ref{SatQevol} demonstrates the qualitative nature of orbital evolution assuming the effective value of $Q$ measured by \cite{lainey:15} is constant in time. Here, we have integrated equation \ref{qdef} backward in time for each moon, finding that the orbital semi-major axis decreases to zero in less than the age of the solar system. In other words, the moons must have formed billions of years after Saturn to be compatible with the theory of a constant tidal $Q$. Although this scenario has been proposed \citep{charnoz:11,crida:12}, we find the resonance locking solution described below to be simpler. A model in which the moons formed well outside of the Roche radius, but still hundreds of millions of years after Saturn \citep{asphaug:13}, remains possible in the resonance locking framework.

In contrast to constant tidal $Q$ models, resonance locking predicts that $t_{\rm tide}$ {\it increases} at small orbital distances (see equation \ref{ttiderl}) where orbital frequencies are higher, for a constant value of $t_\alpha$. The corresponding values of $Q$ (see equation \ref{qdef}) are much larger because of the $Q \propto a^{-13/2}$ scaling for a constant $t_{\rm tide}$. Hence, the effective values of $Q$ for the moons were likely much larger in the past than they are at present, resolving the incompatibility of the small measured values of $Q$ with the age of the solar system. This is demonstrated in Figure \ref{SatQevol}, where we integrate equation \ref{ttiderl} backward in time for each moon, using $t_\alpha = 50 \, {\rm Gyr}$. We caution that the value of $t_\alpha$ may also have been smaller in the past, somewhat offsetting the orbital frequency dependence of equation \ref{ttiderl}, and a detailed orbital evolution should take both these effects into account.

Note also that resonance locking entails the effective values of $Q$ may vary by orders of magnitude over time, depending on whether a given moon is in a resonance lock. Each moon may have spent large amounts of time not involved in resonance locks and migrating outwards on long timescales, until eventually encountering a resonance and migrating outwards more rapidly (or being pushed outward by a MMR). 

A potential problem with resonance locking is that it does not always guarantee convergent orbital migration for two moons. For a constant $t_\alpha$, equation \ref{ttiderl} implies a shorter migration timescale for outer moons. If two moons are both caught in a resonance lock with the same $t_\alpha$, the inner moon will not catch up to the outer moon to establish a MMR. Similarly, if an outer moon is pushed outward by a MMR such that it passes through a mode resonance, it could lock into resonance with the mode and escape the MMR. This could explain how Rhea was able to escape MMRs with other inner moons of Saturn; it simply migrated outward faster. However, the observed MMRs of inner moons require explanation. It is not expected that $t_\alpha$ is exactly the same for all modes, and it is possible that some inner moons have locked into resonance with modes with low values of $t_\alpha$ such that they migrate out faster. Since the mode density of our model is higher near inner moons (see Figure \ref{SatQmodes}), the inner moons may have had more chances to lock with low $t_\alpha$ modes. A more detailed planetary thermal evolution model would be required to investigate this possibility.

\subsection{Tidal Heating}

In Appendix \ref{heat}, we calculate the tidal heating of moons implied by resonance locking. As in conventional tidal theories, the outward migration does not induce tidal heating in a moon unless the moon's eccentricity is increased due to a MMR, leading to heat deposition by eccentricity tides. In this case, the equilibrium tidal heating rate of the inner moon (assuming no tidal migration of the outer moon) is
\beq
\label{eheateq}
\dot{E}_{\rm heat,1} \simeq \frac{1}{j-1} \frac{|E_{2,{\rm orb}}|}{t_{\rm tide}} \, .
\eeq

Using $t_{\rm tide}=35\,{\rm Gyr}$ for Enceladus (see Figure \ref{SatQ}), we calculate its equilibrium heating rate to be $\dot{E}_{\rm heat} \approx 50 \, {\rm GW}$. The actual heating rate may be lower by a factor of a few if we used the measured $Q\sim2000$ found by \cite{lainey:15} (the calculation above corresponded to $Q\sim 1000$), or if we account for outward migration of Dione due to its tidal interaction with Saturn. In any case, resonance locking can account for thermal emission as high as $\dot{E} \approx 16 \, {\rm GW}$ \citep{howett:11}, even if Enceladus is currently in an equilibrium configuration. We note that the heating rate of \citep{howett:11} is controversial, and an updated estimate for heat emitted from Enceladus's tiger stripes alone is 5 GW (\citealt{spencer:13}, see also \citealt{porco:14}). The moon's total radiated power remains unclear, but we predict it may be considerably greater than 5 GW. It is not necessary to invoke the existence of tidal heating cycles to explain Enceladus' observed heat flux, although it remains possible that periodic or outbursting heating events do occur.

For Io, we calculate an equilibrium heating rate of $\dot{E}_{\rm heat} \approx 5\times10^4 \, {\rm GW}$ from equation \ref{edotheat4}, using $t_{\rm tide}=20\,{\rm Gyr}$ (see Figure \ref{JupQ}). This is roughly half the observed heat flux of $10^{5} \, {\rm GW}$ \citep{veeder:94}. The difference may stem from the apparent inward migration of Io due to eccentricity tides \citep{lainey:09}, currently creating moderately enhanced heating compared to the long-term average. Alternatively, the average $t_{\rm tide}$ may be closer to $10\,{\rm Gyr}$, in which case we expect $\dot{E}_{\rm heat} \approx 10^5 \, {\rm GW}$ as observed.

Finally, resonance locking predicts that the tidal heating rate (equation \ref{eheateq}) is only a weak function of semi-major axis. In contrast to a constant tidal $Q$ scenario, we expect the past heating rates of moons like Enceladus and Io to be comparable (within a factor of a few) to the current heating rates, as long as they were in their current MMRs. As in Section \ref{orbits}, we stress that the large current heating rates do not require the moons to have formed after their host planets.

\subsection{Titan and Callisto}

It is possible that Titan and Callisto have experienced significant outward tidal migration despite their larger semi-major axes compared to the inner moons. We posit that Titan and/or Callisto could currently be migrating outward via resonance locking. If so, we predict that $t_{\rm tide} \! \sim \! 2 \, {\rm Gyr}$, and $Q \! \sim \! 20$ for the Saturn-Titan tidal interaction. For the Jupiter-Callisto interaction, we predict $t_{\rm tide} \! \sim \! 2 \, {\rm Gyr}$, and $Q \! \sim \! 1$. Note that values of $Q \! < \! 1$ are possible for migration driven by dynamical tides. A future measurement of such a low $Q$ driving Titan or Callisto's migration would be strong evidence for resonance locking. In this case, these moons may have migrated outward by a significant fraction of their current semi-major axis during the lifetime of the solar system. This migration may have caused Titan to pass through MMRs that excited its eccentricity to its current level \citep{cuk:13}.

Although resonance locking could have occurred for Titan or Callisto in the past, these moons may not currently be in resonance locks. The required mode amplitudes may not be achievable (larger amplitudes are required for moons of larger mass and semi-major axis) due to non-linear effects or resonance saturation.
It is possible that Titan or Callisto were previously in resonance locks, which eventually broke due to the increasingly short migration timescales (and correspondingly large required mode amplitudes) as their orbital frequencies decreased (see equation \ref{ttiderl}).

\section{Conclusions}
\label{conclusions}

We have proposed that a resonance locking process accounts for the rapid outward migration of some of the inner moons of Jupiter and Saturn. During a resonance lock, the outward migration rate is greatly enhanced due to a resonance between the moon's tidal forcing frequency and an oscillation mode of the planet. The oscillation mode could correspond to a gravity mode, or a frequency of enhanced energy dissipation via inertial waves. In either case, the frequency of these oscillation modes change as the planets cool and their internal structures evolve. When a mode frequency crosses a forcing frequency, the moon can be caught in a resonance lock and then migrates outward at a rate comparable to the planet's evolutionary timescale, set by equation \ref{adot}. In some respects, the dynamics of resonance locking is similar to locking into mean motion resonances during convergent orbital migration. 

Resonance locking can only explain the outward migration of Jupiter and Saturn's moons if mode frequencies increase in the rotating frame of the planet, such that their resonant locations move outward (away from the planet), and moons can ``surf" the resonances outward. This will likely be the case for g modes, but the picture is less clear for inertial waves (see Section \ref{inertial}). The occurrence of resonance locking is only weakly dependent on exact mode damping rates and their gravitational coupling with moons, which we find to be amenable to resonance locking (see Appendix \ref{resonancelocks}). Once a resonance lock with a moon is established, the moon's migration rate is determined purely by the outward migration of the resonant location, and is not dependent on the details of the planetary structure, mode frequencies, damping rates, or eigenfunctions (see equation \ref{ttidein}).

Resonance locking predicts an outward migration timescale of order the age of the solar system, as suggested by observations of the Saturn and Jupiter systems (see Figures \ref{SatQ} and \ref{JupQ}, \citealt{lainey:09,lainey:12,lainey:15}). In other words, resonance locking typically predicts low effective tidal quality factors $Q$ governing the migration of moons caught in resonance locks. Resonance locking also predicts different effective values of $Q$ for different moons (via equation \ref{q}), and naturally accounts for the very small value effective $Q$ measured for the migration of Rhea. In the resonance locking paradigm, the outward migration rate $t_{\rm tide}$ is roughly constant for each moon (with some variation, see equation \ref{ttiderl}), rather than the value of $Q$ being constant.

Similarly, resonance locking can account for the large observed heating rates of Io and Enceladus. These high heating rates arise from the short outward migration timescale resulting from resonance locking, resulting in a correspondingly large equilibrium heating rate (equation \ref{eheateq}). Cyclic heating events need not be invoked (except perhaps mild cyclic variation for Io) to account for the current heating, although of course it remains possible that heating cycles do occur.

Finally, resonance locking resolves the problems arising from current-day migration/heating rates, which imply that some moons formed long after their planets if the tidal $Q$ is constant. Instead, resonance locking predicts that the outward migration timescale $t_{\rm tide}$ is nearly constant, such that the effective values of $Q$ were larger in the past. The consequence is that the orbital frequencies of inner moons have likely decreased by a factor of order unity over the lifetime of the solar system (see Figure \ref{SatQevol}), allowing them to have migrated into mean motion resonances with one another, yet still to have formed coevally with Jupiter and Saturn.  We cannot disprove the hypothesis that Saturn's medium-sized inner moons formed after Saturn, but the current rapid migration does not require such a scenario. One possible problem with resonance locking is that it does not guarantee convergent migration of moons. Detailed planetary/tidal evolution models are needed to determine whether it can generally account for the moons' observed mean motion resonances.


\section{Acknowledgments}

We thank Burkhard Militzer, Carolyn Porco, Francis Nimmo, Yanqin Wu, Dong Lai, and Peter Goldreich for useful discussions. JF acknowledges partial support from NSF under grant no. AST-1205732 and through a Lee DuBridge Fellowship at Caltech. JL is supported by TAC and CIPS at UC Berkeley. EQ was supported in part by a Simons Investigator award from the Simons Foundation and the David and Lucile Packard Foundation.

\bibliography{SatBib}

\appendix

\section{Oscillation Modes and Tidal Dissipation}
\label{modes}

\subsection{Gravito-inertial Modes}

Here we describe our method of calculating Saturn's oscillation modes and their effect on tidal dissipation. We adopt the Saturn model shown in Figure 2 of \cite{fuller:14}, which reproduces Saturn's mass, radius, spin frequency, and gravitational moment $J_2$. The most important feature of this model is that it contains a stably stratified region outside the core, at radii $0.1 \lesssim r/R \lesssim 0.4$, which supports the existence of gravity modes. This particular model has a Brunt-V\"{a}is\"{a}l\"{a} frequency significantly larger than the tidal forcing frequencies, $N \sim 8 \omega_{\rm f}$, such that resonantly forced g modes approximately obey WKB relations.

To calculate the oscillation modes of this model, we adopt the traditional approximation (see e.g., \citealt{bildsten:96,lee:97}) to find the so-called Hough modes on the low $\ell$ mode branches. In stably stratified regions, the traditional approximation is valid because $N>\omega_{\rm f}$ and we find that the resonant modes have horizontal displacements larger than radial displacements, $\xi_{\perp} > \xi_r$. The Hough modes calculated from the traditional approximation are composed of many spherical harmonic degrees $\ell$. However, each branch of modes can be traced back to a single spherical harmonic degree $\ell$ in the non-rotating limit, which we will refer to as as the $\ell$ of each mode. The adiabatic mode frequencies and eigenfunctions are calculated with standard numerical methods (see \citealt{fuller:14b} for more detailed description) and with the usual reflective boundary conditions. We normalize the modes via (c.f. \citealt{schenk:02,lai:06}) the orthonomality condition
\beq
\label{modeorthrot}
\langle \xi_\alpha | \xi_\beta \rangle = \bigg[ \delta_{\alpha \beta} - \frac{2}{\omega_\alpha + \omega_\beta} W_{\alpha \beta} \bigg] M_{\rm p} R_{\rm p}^2 \, 
\eeq
where
\beq
\label{modeorth}
\langle \xi_\alpha | \xi_\beta \rangle \equiv \int dV \rho \bxi_\alpha^* \cdot \bxi_\beta 
\eeq
and 
\beq
\label{Wdef}
W_{\alpha \beta} \equiv \int dV \rho \bxi_\alpha^* \cdot \Big( i {\bf \Omega_s} \times \bxi_\beta \Big) \, .
\eeq

For our purposes, we calculate only the retrograde modes along the g mode branches (i.e., no Rossby modes) that are symmetric across the planetary equator. We have calculated $(\ell,m)$ combinations of $(2,2)$, $(3,1)$, $(3,3)$, $(4,2)$, and $(4,4)$ which generally couple most strongly to the tidal potential of the perturber. We have restricted our calculations to include only low/medium order g modes and fundamental modes whose frequencies are somewhat near $\omega \sim m \Omega_{\rm p}$ and can be resonantly excited.

We note that many of the relevant oscillation modes have $\omega_\alpha \! < \! 2 \Omega_s$ and therefore lie within the sub-inertial regime in which inertial waves may propagate in the convective envelope. This could significantly change the character of these oscillation modes, especially their eigenfunction within the convective envelope, a fact which has been explored in some previous works (e.g., \citealt{dintrans:99,dintrans:00,mirouh:15}). In fact, the mode eigenspectrum is undoubtedly more complex than the one shown in Figure \ref{SatQmodes}, containing more dips in $Q$ associated with tidal dissipation via gravito-inertial waves focused onto wave attractors or absorbed at critical layers (i.e., irregular modes). Were we to require a precise determination of mode frequencies and eigenfunctions, our technique would not be viable. However, for our purposes what matters is that modes exist in this frequency range (both regular and irregular modes would suffice for resonance locking), and that we can roughly estimate their gravitational potential perturbations and damping rates. Although our computed eigenfunctions and their potential perturbations may be quantitatively inaccurate, this is unimportant unless it prevents the resonance locking process from occurring (see Section \ref{reslocking}).

Exact mode damping rates are also difficult to calculate. However, in a resonance lock the tidal dissipation rate is {\it independent} of the mode damping rates. The mode damping rates affect only the ability of the system to enter into and maintain a resonance lock. As long as realistic damping rates are within a few orders of magnitude of our estimated damping rates, the basic picture we advance may still occur.

\subsection{Tidal Energy Dissipation}

Once the modes have been obtained, we calculate their forced amplitude in a manner similar to \cite{schenk:02,lai:06}. In the adiabatic limit (i.e., the mode amplitude changes slowly), the mode's forced amplitude is 
\beq
\label{modeamp}
a_\alpha = \frac{1}{M_{\rm p} R_{\rm p}^2} \frac{\langle \bxi_\alpha | - \bnab U \rangle}{2 \omega_\alpha (\omega_\alpha - \omega_{\rm f} - i \gamma_\alpha)} \,.
\eeq
Here, $\bxi_\alpha$ is the mode displacement eigenfunction, $U$ is the tidal potential of the companion, and $\omega_{\rm f}$ is the tidal forcing frequency defined on the right hand side of equation \ref{res1}. Defining the dimensionless tidal overlap integral
\beq
\label{Qalpha}
Q_{\alpha,\ell,m} = \frac{G}{R_{\rm p}^{\ell+3}} \frac{\langle \bxi_\alpha | \bnab (r^\ell Y_{\ell,m}) \rangle}{\omega_\alpha^2}   \, ,
\eeq
the mode amplitude is
\beq
\label{modeamp2}
a_\alpha = \frac{1}{2} \frac{\omega_\alpha}{\omega_\alpha-\omega_{\rm f} - i \gamma_\alpha} \sum_{\ell} \epsilon_{\ell,m} Q_{\alpha,\ell,m} \, ,
\eeq
with the dimensionless tidal amplitude
\beq
\label{epslm}
\epsilon_{\ell,m} = W_{\ell,m} \frac{M_{\rm m}}{M_{\rm p}} \bigg( \frac{R_{\rm p}}{a_{\rm m}} \bigg)^{\ell+1} \, ,
\eeq
with $W_{\ell,m}$ a constant of order unity defined in equation 2 of \cite{lai:06}. The sum in equation \ref{modeamp2} results from the fact that each mode is a superposition of multiple spherical harmonic degrees $\ell$ and thus couples to multiple components of the tidal potential. For simplicity we only include low degrees $\ell \leq 4$ in our sums because low $\ell$ components dominate the tidal coupling. The tidal energy dissipation rate associated with the mode amplitude in equation \ref{modeamp2} is
\beq
\label{edamp}
\dot{E}_{\alpha} = \bigg[\sum_{\ell} \epsilon_{\ell,m} Q_{\alpha,\ell,m} \bigg]^2 \frac{\omega_\alpha^2}{(\omega_\alpha-\omega_{\rm f})^2 + \gamma_\alpha^2} \omega_{\rm f}^2 \gamma_\alpha M_{\rm p} R_{\rm p}^2 \, .
\eeq

Equation \ref{edamp} is the tidal energy dissipation measured in the rotating frame of the planet. In the inertial frame, the orbital energy gained by the moons is
\beq
\label{etide}
\dot{E}_{\rm tide,\alpha} = \frac{m \Omega_{\rm m}}{\omega_{\rm f}} \dot{E}_{\alpha} \, .  
\eeq
The total tidal energy transfer rate is the sum of that provided by each mode,
\beq
\label{etidetot}
\dot{E}_{\rm tide} = \sum_\alpha \dot{E}_{\rm tide,\alpha} \, .
\eeq
Near resonance, the resonant mode typically dominates the total tidal energy dissipation rate. The corresponding angular momentum transfer rate is $\dot{J}_{\rm tide,\alpha} = \dot{E}_{\rm tide,\alpha}/\Omega_{\rm m}$. Moons therefore maintain circular orbits as they migrate outward, unless their eccentricity is excited by a MMR with another moon.

\subsection{Mode Damping Rates}
\label{damp}

Our mode calculations are adiabatic and do not self-consistently calculate the mode damping rates. Here we calculate order-of-magnitude estimates for the mode damping rates. We find that thermal diffusion is unlikely to be important for g-modes trapped deep in the planet, and that mode damping is likely dominated by an effective convective viscosity. 

The mode damping rate due to convective viscosity is
\beq
\label{convdamp}
\gamma_\alpha \approx \frac{\int d M_{\rm con} \nu_{\rm con} k_r^2 \bxi^* \cdot \bxi}{\int d M\bxi^* \cdot \bxi} \, .
\eeq
Here, the integral in the numerator is taken over regions of the planet which are convective, $k_r$ is the radial wavenumber of the mode, and $\nu_{\rm con}$ is the effective convective viscosity. 

A first guess for the effective convective viscosity is 
\beq
\label{nuconv1}
\nu_{\rm con} \sim l_{\rm con} v_{\rm con} \, ,
\eeq
where $l_{\rm con}$ and $v_{\rm con}$ are the convective mixing length and velocity. However, this effective viscosity is reduced if the mode period is shorter than the convective turnover time $t_{\rm con} = l_{\rm con}/v_{\rm con}$ (see \citealt{ogilvie:12} for a recent discussion). The amount of suppression is still debated, but we adopt the prescription of \cite{goldreich:77}, in which the effective convective viscosity is
\beq
\label{nuconv2}
\nu_{\rm con} \sim l_{\rm con} v_{\rm con} {\rm min} \big[ 1, (t_{\rm f}/t_{\rm con})^2 \big] \, .
\eeq
Here, $t_{\rm f} = \omega_{\rm f}^{-1}$ is the tidal forcing time associated with a given moon. 

We find typical mode lifetimes $t_{{\rm damp},\alpha} = \gamma_\alpha^{-1}$ (evaluated using equation \ref{nuconv2}) of $10^9 \, {\rm s} \lesssim t_{{\rm damp},\alpha} \lesssim 3 \times 10^{11} \, {\rm s}$. Somewhat surprisingly, we find mode lifetimes are largest for higher order g modes and smallest for f modes. The reason is that high-order g modes are trapped in radiative regions (where they cannot be convectively damped), whereas f modes (and low-order g modes to a lesser extent) have more inertia in convective regions and suffer more convective damping. 

When we calculate mode damping rates using the simple prescription of equation \ref{nuconv1}, we find typical mode lifetimes of $1.5 \times 10^8 \, {\rm s} \lesssim t_{{\rm damp},\alpha} \lesssim 3 \times 10^{9} \, {\rm s}$. We consider these to be upper limits to plausible mode damping rates, whereas damping rates calculated using the prescription of equation \ref{nuconv2} are lower limits. Using the intermediate prescription of \cite{zahn:66} yields damping rates in between those listed above.

The important feature of the damping rates we estimate is that they lie in the regime $t_{\rm f} \ll t_{{\rm damp},\alpha} \ll t_{\rm evol}$. Therefore, all modes can be regarded as weakly damped, yet damped strongly enough that mode amplitudes can be reliably calculated using the adiabatic approximation (see below).

\section{Ability to Sustain Resonance Locks}
\label{resonancelocks}

Resonance locking can proceed if several conditions (outlined in detail in \citealt{burkart:14}) are satisfied. First, tides must operate in the weak damping limit (equation 32 of \citealt{burkart:14}), i.e., tidal dissipation must be greatly enhanced near resonances so that $t_{\rm tide} \! < \! t_{\rm evol}$ in Figure \ref{SatQmodes} only near resonances. We find this is indeed the case for our model. If there is an additional source of tidal dissipation such as viscoelasticity in a solid core, this could lower the equilibrium tide value of $Q$ such that the weak damping limit does not apply for the inner moons. In this case, resonance locking could not operate for the inner moons, but it could still work for outer moons.

Second, resonance locking can only occur if the fixed point in Figure \ref{SatQmodes} is a stable fixed point, which arises if the motion of the resonant location in Figure \ref{SatQmodes} is in the same direction as the direction of moon migration, i.e., away from the planet. This requires the mode frequencies to be decreasing in an inertial frame such that they continue to resonate with the decreasing orbital frequency. However, the mode frequencies measured in the rotating frame of the planet (equation \ref{res1}) must be {\it increasing}, i.e., $t_\alpha > 0$. In the language of \cite{burkart:14}, this is equivalent to the requirement $\Gamma_{\rm dr} < 0$. In Section \ref{inertial}, we showed this may not always occur for resonances with inertial waves. For gravity modes, the dispersion relation (in the non-rotating limit) results in 
\beq
\label{gmode}
t_\alpha = \frac{\omega_\alpha}{\dot{\omega}_\alpha} = \frac{N}{\dot{N}} \, .
\eeq
For a planet that is building a stably stratified region through helium sedimentation or core erosion, we expect the buoyancy frequency $N$ to be increasing such that $t_\alpha$ is positive and resonance locking can occur. Including rotation allows for the existence of gravito-inertial modes whose dispersion relation is more complicated. The modes may also undergo avoided crossings as the planet cools, giving rise to many possible values of $t_\alpha$, although we expect that $t_\alpha$ will typically be positive if $\dot{N}$ is positive, and we expect $t_\alpha \sim t_{\rm cool}$ to order of magnitude.  

In this paper, we have used the ``adiabatic" limit to calculate mode amplitudes, which is valid when (see equation 46 of \citealt{burkart:14})
\beq
(\omega_{\rm f} - \omega_\alpha)^2 + \gamma_\alpha^2 > \omega_\alpha/t_\alpha \, .
\eeq
For the fixed points shown in Figure \ref{SatQmodes}, the adiabatic limit is valid. However, we note that the modes do not always satisfy $\gamma_\alpha^2 > \omega_\alpha/t_\alpha$ for $t_\alpha = 50 \, {\rm Gyr}$. Therefore, the adiabatic limit may not be valid for more massive moons such as Titan, given the values of $\gamma_\alpha$ calculated above. Damping rates larger by 1-2 orders of magnitude (which is possible since the damping rates above should be regarded as lower limits) would guarantee adiabaticity for any moon. Damping rates larger by more than 3-4 orders of magnitude (unlikely) would eliminate the existence of the stable fixed points (i.e., resonance locking could not occur before the resonances saturate). 

Additionally, we find that resonances satisfy the corresponding condition for stable resonance locking (equation 50 of \citealt{burkart:14}). Larger damping rates would strengthen this result. We conclude that resonance locking can occur in giant planet systems as long as the resonant locations of some modes evolve away from the planet, as shown in Figure \ref{SatQmodes}.

\section{Tidal Heating of Moons}
\label{heat}

Moons caught in MMR will excite each other's eccentricity and/or inclination. This will allow for tidal dissipation within each moon, generating tidal heating of the moons. Assuming that tidal dissipation within the moons keeps their orbits nearly circular, the rate of change of the moons' orbital energy due to outward migration is given by equation \ref{edottide}. However, the net change in the moons' orbital energy is
\begin{align}
\label{edotorb}
\dot{E}_{\rm orb} &= \dot{E}_{1,{\rm orb}} + \dot{E}_{2,{\rm orb}} \nonumber \\
&= -\frac{\dot{a}_1}{a_1} E_{1,{\rm orb}} - \frac{\dot{a}_2}{a_2} E_{2,{\rm orb}} \nonumber \\
&= -\frac{\dot{a}_1}{a_1} \big( E_{1,{\rm orb}} + E_{2,{\rm orb}} \big) \, .
\end{align}
The last line follows from the MMR condition of equation \ref{a12}. The energy which must be dissipated as heat is
\beq
\label{edotheat}
\dot{E}_{\rm heat} = \dot{E}_{\rm tide} - \dot{E}_{\rm orb} \, .
\eeq

In general, the heating rate depends on the relative rates of outward migration due to tides in the planet raised by each moon. Let us assume once again that the outward migration of the moons is caused solely by the tidal interaction of the inner moon with the planet. In this case, $\dot{E}_{\rm tide} = \dot{E}_{1,{\rm tide}} = \Omega_1 \dot{J}_{1,{\rm tide}} = \Omega_1 \dot{J}_{\rm orb}$, and using equation \ref{jdot} we have
\beq
\label{edotheat2}
\dot{E}_{\rm heat} = \Omega_1 \bigg[ \frac{1}{2} \frac{\dot{a}_1}{a_1} \big(J_1 + J_2\big) \bigg] + \frac{\dot{a}_1}{a_1} \big( E_{1,{\rm orb}} + E_{2,{\rm orb}} \big) \, .
\eeq
Using $E_{1,{\rm orb}} \simeq - \Omega_1 J_1/2$ and likewise for moon 2, we have 
\begin{align}
\label{edotheat3}
\dot{E}_{\rm heat} & \simeq \frac{1}{2} \frac{\dot{a}_1}{a_1} \big( \Omega_1 - \Omega_2 \big) J_{2,{\rm orb}} \nonumber \\
& \simeq \frac{1}{j-1} \frac{|E_{2,{\rm orb}}|}{t_{\rm tide}} \, .
\end{align}
For small eccentricities, this is identical to (but simpler than) equation 15 of \cite{meyer:07} and equation 7 of \cite{lissauer:84} (with $m_3$ set to zero). The tidal heat dissipated within moon 1 depends on its outward migration rate and the mass of moon 2. For three moons in a resonant chain, the result is
\begin{align}
\label{edotheat4}
\dot{E}_{\rm heat} \simeq \frac{1}{t_{\rm tide}} \bigg[ \bigg(\frac{\Omega_1}{\Omega_2} -1 \bigg)|E_{2 {\rm ,orb}}| + \bigg(\frac{\Omega_1}{\Omega_3} -1 \bigg)|E_{3,{\rm orb}}| \bigg] \, .
\end{align}

As in \cite{meyer:07}, we calculate the equilibrium eccentricity at which equation \ref{edotheat3} is equal to the tidal heating rate of a moon due to its eccentricity,
\beq
\label{eheat}
\dot{E}_{\rm heat} = \frac{21}{2} \frac{k_1}{Q_1} \frac{G M_{\rm p}^2 R_{1}^5}{a_1^6} \Omega_1 e_1^2 \, .
\eeq
Here, $k_1$ and $Q_1$ refer to the Love number and tidal quality factor of the moon, and $e_1$ is its eccentricity. In the limit that the outward migration is driven solely by tidal torques on moon 1, the equilibrium eccentricity is
\beq
\label{eeq}
e^2_{\rm eq} = \frac{1}{7(j-1)} \frac{M_1 M_2}{M^2} \bigg(\frac{R_{\rm p}}{R_1}\bigg)^5 \frac{Q_1}{Q_{{\rm p},1}} \frac{k_{\rm p}}{k_1} \, ,
\eeq
where we have denoted the effective tidal quality factor driving the outward migration of moon 1 as $Q_{{\rm p},1}$. 


Equations \ref{edotheat3} and \ref{edotheat4} may be good approximations for Enceladus and Io, which are caught in eccentric MMRs with outer moons. However, it is likely not valid for Mimas, which is caught in an inclination MMR with Tethys. In this case, the inclinations of Mimas and Tethys are excited as they migrate outward, although this inclination likely cannot be damped out via tidal dissipation in Mimas \citep{luan:14}. However, it may be possible that the orbital inclination can be damped out through interactions between Mimas and Saturn's rings.

\end{document}